\documentclass[10pt]{wlscirep}
\usepackage{amsmath,graphicx,amssymb,braket,xcolor,subfigure,upgreek}
\usepackage{wrapfig}
\usepackage{color}

\title{Selective protected state preparation of coupled dissipative quantum emitters}

\author[1]{D. Plankensteiner}
\author[1]{L. Ostermann}
\author[1]{H. Ritsch}
\author[1,*]{C. Genes}
\affil[1]{Institut f\"ur Theoretische Physik, Universit\"at
Innsbruck, Technikerstrasse 21a, A-6020 Innsbruck, Austria}

\affil[*]{claudiu.genes@uibk.ac.at}

\begin{abstract}
Inherent binary or collective interactions in ensembles of quantum emitters induce a spread in the energy and lifetime of their eigenstates. While this typically causes fast decay and dephasing, in many cases certain special entangled collective states with minimal decay can be found, which possess ideal properties for spectroscopy, precision measurements or information storage. We show that for a specific choice of laser frequency, power and geometry or a suitable configuration of control fields one can efficiently prepare these states. We demonstrate this by studying preparation schemes for strongly subradiant entangled states of a chain of dipole-dipole coupled emitters. The prepared state fidelity and its entanglement depth is further improved via spatial excitation phase engineering or tailored magnetic fields.
\end{abstract}

\begin{document}
\thispagestyle{empty}

\maketitle

\section*{Introduction}
\indent Ensembles of effective two-level quantum emitters consisting
of single atoms, ions, or defects in solids are employed
ubiquitously in quantum optics and quantum
information~\cite{lukin2003colloquium}. They are the basis for
precision spectroscopy or atomic clock setups, as well as for
experiments testing fundamental concepts of quantum physics or
implementations of the strong coupling cavity QED (quantum
electrodynamics) regime~\cite{lukin2001dipole,hammerer2010quantum}.
In the absence of direct particle-particle interactions, larger
ensembles allow for faster, more precise
measurements~\cite{wineland1992spin} via a  scaling of the effective
single photon to matter coupling strength $g$ by a factor $\sqrt{N}$
(with system size $N$) and a reduction of the quantum projection
noise (by $1/\sqrt{N}$)~\cite{sangouard2011quantum,leroux2010focus}.

For any precise measurement one has to externally prepare, control
and measure the particle dynamics. Hence, the emitters are almost
unavoidably coupled to their environment. A suitable theoretical
framework to model such experiments is open system dynamics with a
coupling to a fluctuating thermal bath. At optical frequencies this
can often be approximated by the zero effective temperature
electromagnetic vacuum
field~\cite{davies1976quantum,gardiner2004quantum}. Still, extra
perturbations by a thermal environment and background gas collisions
cannot be avoided.

In a laboratory experiment the particles need to be confined in a
finite spatial volume that can be addressed by laser beams. Thus,
increasing particle numbers will lead to higher densities, where
direct particle-particle interactions as well as environmentally
induced collective decoherence can no longer be neglected. For
optical transition frequencies a critical density is conventionally
assumed at the point where the average particle separation is of the
order of an optical wavelength~\cite{gheri1999quantum}. Above this
limit vacuum fluctuations tend to become uncorrelated and decay
becomes independent. However, recent calculations have shown that
collective states can exhibit superradiance and subradiance even at
much larger distances~\cite{zoubi2010metastability} as long as the
bandwidth of the emission is small enough.

\noindent In many typical configurations and in optical lattices in
particular, the particle-particle interaction is dominated by binary
dipole-dipole couplings, with its real part inducing energy shifts
and its imaginary part being responsible for collective
decay~\cite{lehmberg1970radiation,freedhoff1979collective}.
Generally, this interaction is associated with dephasing and decay.
However, recently it has been found that under special conditions
also the opposite can be the case and these interactions can lead to
a synchronization ~\cite{zhu2014quantum} or even a blockade of the
decay~\cite{zoubi2008bright}.

\noindent Oftentimes it is assumed that while such states exist,
they cannot be prepared by lasers as they are strongly decoupled
from the radiation fields. However, it was recently proposed that
individual instead of overall addressing of the atoms can push the
many particle system to evolve towards subspaces protected from
decay or dephasing~\cite{ostermann2013protected}. When applied to
Ramsey spectroscopy such states have been shown to exhibit
frequency sensitivities superior even to those obtained from
non-interacting ensembles~\cite{ostermann2014protected}. However,
apart from special cases with an optimal lattice size and excitation
angle, it is not so obvious how to implement such precise a control.

\begin{figure}[t]
\includegraphics[width=\textwidth]{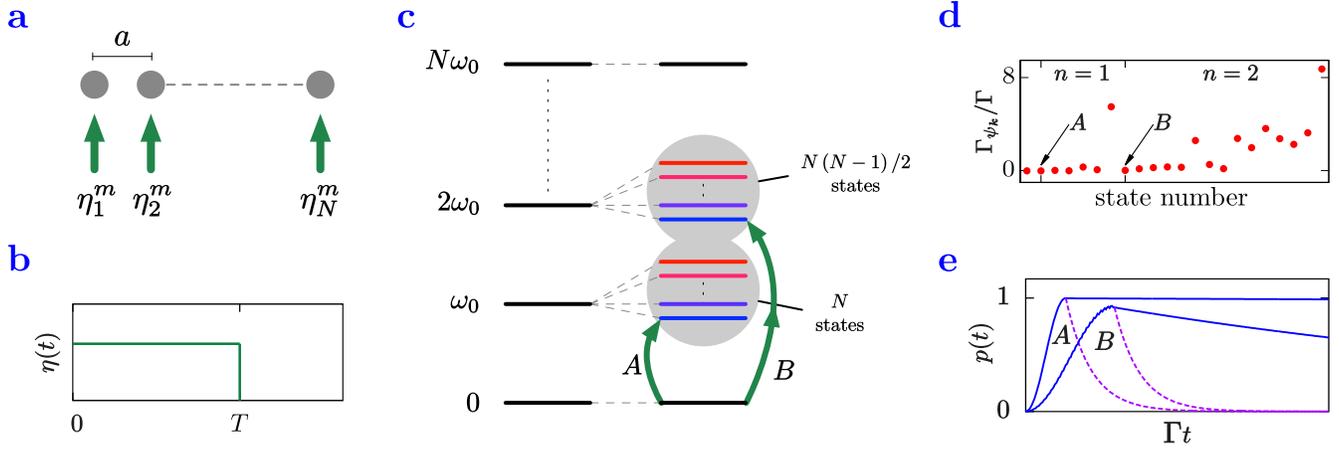}
\caption{\emph{Selective state preparation procedure}. \textcolor{blue}{a)} A chain of $N$ closely spaced quantum emitters (separation $a$ with $k a \ll 1$, $k$ being the laser wave number) are individually driven with a set of pumps $\{\eta_j^m\}$. \textcolor{blue}{b)} The lasers are turned on for a time $T$, optimized such that an effective $\pi$-pulse into the desired subradiant target state is achieved. \textcolor{blue}{c)} Level structure for the $N$ systems where the $C_n^N$-fold degeneracy of a given $n$-excitation manifold is lifted by the dipole-dipole interactions. The target states are then reached by energy resolution (adjusting the laser frequency) and symmetry (choosing the proper $m$). \textcolor{blue}{d)} Scaling of the decay rates of energetically ordered collective states starting from the ground state (state index $1$) up to the single- and double-excitation manifolds for $6$ particles at a distance of $a=0.02\,\lambda_0$. The arrows identify the decay rates for the lowest energy states in the single (A) and double (B) excitation manifolds. \textcolor{blue}{e)} Numerical results of the time evolution of the target state population for $N=6$ and $a=0.02\,\lambda_0$ during and after the excitation pulse. Near unity population is achieved  for both example states A (where we used $\eta=0.53\,\Gamma$) and B (for $\eta=2.44\,\Gamma$) followed by a subradiant evolution after the  pulse time $T$ shown in contrast to the independent decay with a rate $\Gamma$ (dashed line).}
\label{fig1}
\end{figure}

In this work we highlight the surprising fact that interaction
induced level shifts can be used to aid in preparing such states. In
many cases the magnitude of the shifts a state experiences and its
lifetime are tightly connected allowing one to identify and address
interesting states via energy resolution. As a generic ensemble we
particularize to a 1D regular chain of quantum emitters coupled by
dipole-dipole interactions with a tunable magnitude (by varying the
interparticle separation). Collective coupling to the vacuum
leads to the occurrence of subradiant as well as superradiant
excitonic states~\cite{zoubi2010metastability}. In particular, the
subradiant states should prove extremely useful for quantum
information as well as metrology applications as they exhibit
robust, multipartite quantum correlations. As mentioned above, the
atoms' interactions provide a first handle for target state
selection as they lead to energy resolved collective states.
Furthermore, using a narrow bandwidth laser excitation matched to
the target states both in energy and symmetry allows for a selective
population transfer from the ground state via an effective Rabi
$\pi$-pulse.

\noindent In many cases, however, the required phase structure of
the target state is not compatible with the excitation laser phase
so that only a very weak coupling can be achieved. On the other
hand, increasing the laser power reduces spectral selectivity by an
unwanted addressing of off-resonant but strongly coupled states.
Hence, to address a larger range of states of practical interest, we
also propose and analytically study new methods of phase imprinting
via a weak spatial magnetic field gradient. The small relative phase
shifts increase the effective coupling to groups of emitters via a
nonuniform phase distribution. With this method any state may
acquire a finite laser coupling to the ground state via the
magnetically induced level shifts resulting in an efficient
population transfer with a minimal compromise on lifetime.

The considered setup is a chain (see
Fig.~\ref{fig1}\textcolor{blue}{a}) of $N$ identical two-level
systems (TLS) with levels $\ket{g}$ and $\ket{e}$ separated by a
frequency of $\omega_0 $ (transition wavelength $\lambda_0$) in a
geometry defined by the position vectors $\left \lbrace \mathbf{r}_i
\right \rbrace$ for $i=1,...N$. For each $i$, operations on the
corresponding two-dimensional Hilbert space are written in terms of
the Pauli matrices $\sigma_i^{x,y,z}$ and raising/lowering operators
$ \sigma _i^\pm$ connected via $\sigma_i^x = \sigma_i^+ + \sigma_i^-
$, $\sigma_i^y =-i(\sigma_i^+ -\sigma_i^-)$ and $\sigma_i^z
=\sigma_i^+\sigma_i^- -\sigma_i^-\sigma_i^+$. The complete
Hamiltonian describing the coherent dynamics is
\begin{align} \label{H}
{H} &= {H}_0+{H}_{dip} = \omega_0\sum_i\sigma_i^+\sigma_i^- + \sum_{i\neq j}\Omega_{ij}\sigma_i^+\sigma_j^-,
\end{align}
where ${H}_0$ is the free Hamiltonian and has degenerate energy
levels (degeneracy $C_n^N=N!/(N-n)!n!$ for level $n$) ranging from
$0$ for the ground state to $N\omega_0$ for the highest excited
state. The second term ${H}_{dip}$ describes interactions between
pairs of TLS which can be induced either by an engineered bath (such
as a common, fast evolving optical cavity field) or by the inherent
electromagnetic vacuum. We denote the couplings between emitters $i$
and $j$ by $\Omega_{ij}$ and particularize to the case of a
free-space one dimensional equidistant chain of TLS with small
interparticle distances $a$ such that $a\ll\lambda_0$ (as depicted
in Fig.~\ref{fig1}\textcolor{blue}{a}).

For the sake of simplicity, we use dipole moments perpendicular to
the chain for all numerical computations. To a good approximation,
in the limit of $k_0a\ll1$, the nearest-neighbor (NN) assumption can
be used (such that $\Omega _{ij}=\Omega \delta_{ij\pm1}$) and exact
solutions in the single-excitation manifold can be found
\cite{zoubi2012optical}. Within this subspace and approximation, the
Hamiltonian assumes the form of a tridiagonal symmetric Toeplitz
matrix with $\omega_0$ on the diagonal and $\Omega$ above and below
the diagonal. The solutions are readily available~\cite{bottcher2012introduction}
with eigenvalues $\omega_0+\epsilon_m$ for an index $m$ running from $1$
to $N$, where $\epsilon_m =2\Omega\cos{[\pi m/ (N+1)]}$ are the dipole-induced energy shifts. The corresponding eigenstates of the Hamiltonian are then
\begin{align}
\ket{m} = \sum_{j} f^{m}_j\sigma _j^+\ket{G},~\text{with }f^{m}_j = \sqrt{\frac{2}{N+1}} \sin{\left(\frac{\pi m j}{N+1}\right)},
\end{align}
where we used $\ket{G}=\ket{g}^{\otimes N}$.

Spontaneous decay via a coupling to the free radiation modes in the
evolution of the system can be included in a generalized Lindblad
form~\cite{gardiner2004quantum},
\begin{equation} \label{L_def}
\mathcal{L}[\rho ]= \frac{1}{2} \sum_{i,j} \gamma _{ij} \left( 2 \sigma_i^- \rho \, \sigma_j^+ -\sigma_i^+ \sigma_j^- \rho -\rho \, \sigma_i^+ \sigma_j^- \right),
\end{equation}
where the $\gamma_{ij}$ denote collective damping rates arising from
the coupling to a common radiation field. These rates also strongly
depend on the atomic distances $a$ with two prominent limiting cases
of $\gamma _{ij}(a\rightarrow\infty)=\Gamma \delta_{ij}$
(independent emitters limit) and $\gamma _{ij}(a\rightarrow
0)=\Gamma$ (the Dicke limit \cite{dicke1954coherence}). In general,
one can perform a transformation of the Liouvillian into a new basis
by diagonalizing the $\gamma_{ij}$ matrix. This procedure leads to a
decomposition into $N$ independent decay channels with both
superradiant ($> \Gamma$) and subradiant (robust) decay rates
($<\Gamma$)~\cite{ostermann2014protected}. Note, however, that the
states corresponding to these channels generally do not coincide
with energy eigenstates of the Hamiltonian, so that we cannot reduce
the system dynamics to simple rate equations.

\section*{Results}

\subsection*{Selective state preparation}

\noindent \textbf{Tailored coherent excitation}. As mentioned above,
our dipole coupled systems possess states with a large range of
radiative lifetimes and energy shifts. Depending on the desired
application particular states can be highly preferable over others.
In a first straightforward approach we now illustrate that in
principle it is possible to access a desired collective state simply
by a selective coherent driving with a properly chosen amplitude and
phase for each TLS. This is described by the Hamiltonian
\begin{align} \label{H_eta}
{H}_m  =  \sum_j \eta_j^m(\sigma_j^+ e^{-i \omega_l t} + \sigma_j^- e^{i \omega_l t}),
\end{align}
with a suitably chosen set of $\eta_j^m$. For a targeted eigenstate
in the single-excitation manifold, some analytical insight on how to
choose these amplitudes can be gathered from the state's symmetry.
For energy eigenstates this can be found quite reliably within the
NN approximation~\cite{zoubi2013excitons}. In an equidistant finite
chain our calculation suggests the following choice of driving fields at laser
frequency $\omega_l$,
\begin{align}
\eta_j^m=\eta\sin\left(\frac{\pi m j}{N+1}\right),
\end{align}
chosen to fit the symmetry of a target state $\left| m \right \rangle$.

The selectivity of the excitation process can be further improved by
an \textit{energetically resolved excitation} of a given state
$\left| m \right \rangle$ by a proper choice of the laser frequency
$\omega_l=\omega_0+\epsilon_m$ and its bandwidth. This is possible
due to the interaction induced level splitting from ${H}_{dip}$ (as
depicted in Fig.~\ref{fig1}\textcolor{blue}{c}). Indeed, in
perturbation theory and in a frame rotating at $\omega_l$ the
evolution of the system starting from the ground state  up to a
normalization factor leads to
\begin{equation}
e^{-i {H}_m t} \ket{G}\simeq \ket{G}- i\eta t \ket{m}.
\end{equation}
The success of the corresponding process is illustrated in the
sequence of plots in Fig.~\ref{fig1}, where the $\ket{m=N}$ state
with $n=1$ is considered (target state A) and accessed via the
combination $\eta^N_j$ of pumps lasting for a duration $T$.

Numerical simulations were performed on a six-atom chain with
driving strength $\eta=0.53\,\Gamma$ at an interatomic separation of
$a=0.02\,\lambda_0$. The time for which the pumps are switched on is
$T=1.58\,\Gamma^{-1}$ which is considerably shorter than the time
scale governed by the decay rate of $0.0009\,\Gamma$ of the target
state. The resulting dynamic is an effective $\pi$-pulse (efficiency
of $99.94\% $) flipping the population into the state $\ket{m=N}$
followed by an extremely slow decay, indicating the robustness of
the target state (as seen in curve A of
Fig.~\ref{fig1}\textcolor{blue}{e}).

It is, of course, desirable to target higher excitation manifolds as
well. In the absence of analytical expressions or good
approximations for the target states, we employ phases that yield
maximal asymmetry, i.e. $\bar\eta_j=\eta(-1)^j$ for any $j=1,...,N$.
Such a driving can be expected to address collective states, where
the fields emitted by any two neighboring particles interfere
destructively~\cite{zoubi2008bright} (similar to a previously investigated mechanism~\cite{ostermann2013protected}). Numerical simulations
show that the resulting collective states indeed exhibit the lowest
energy shifts of the targeted manifold and can be expected to be
long lived. The resonance condition for a specific state
$\ket{\psi}$ within the manifold $n$ is
$n\omega_l=n\omega_0+\delta\omega_\psi$, where
$\delta\omega_\psi=\bra{\psi}{H}_{dip}\ket{\psi}$. As an
illustration, the curve B in Fig.~\ref{fig1}\textcolor{blue}{e}
shows an almost perfect efficiency ($98.36\%$) two-photon
$\pi$-pulse allowing for a population transfer to the longest-lived
collective state in the second excitation manifold of $N=6$ emitters
separated by $a=0.02\,\lambda_0$. The chain was driven with a
strength of $\eta=2.44\,\Gamma$ for a time $T=3.44\,\Gamma^{-1}$,
which again is significantly shorter than the natural time scale
given by the target state decay rate of $0.0402\,\Gamma$.

Let us add a comment on the practical implementation of such an
addressing. In typical current experimental configurations for
clocks based on 1D magic wavelength
lattices~\cite{bloom2014optical,ushijima2015cryogenic} the atoms are
very close and hardly allow for an individual direct particle
addressing. One is largely limited by a quasi plane wave driving,
which typically addresses all particles with equal intensity. If the
pump light is applied perpendicularly to the trap, the evolution is
governed by a symmetric Hamiltonian ${H}_{sym}$, obtained from
equation~(\ref{H_eta}) with an equal pump amplitude $\eta^m_j=\eta$
for any $m$ and $j$. A laser excitation from the ground state into
the state $\ket{m}$ is connected to the coupling amplitude
$\chi_m=\bra{m}{H}_{sym}\ket{G}=\eta\sum_i f_i^m$, which yields
\begin{align}
\chi_m = \left\{ \begin{array}{l l} 0 & \quad \text{if $m$ is even,}\\
\frac{\sqrt{2}\eta}{\sqrt{N+1}}\cot{\left(\frac{m\pi}{2N+2}\right)} & \quad \text{if $m$ is odd.}
\end{array} \right.
\end{align}

\noindent We will refer to states with even $m$ as \textit{dark
states} as they cannot be accessed by the laser excitation and call
the remaining ones \textit{bright states} \cite{zoubi2008bright}. In
the limit of large atom numbers $N\gg 1$, it is of interest to
investigate the two cases, where $m\ll N$ and $m\sim N$, for states
at the top/bottom of the manifold. In the first case, the function
for the driving yields $\chi_m\approx \eta\sqrt{8N}/m\pi$, whereas
in the other case we have $\chi_m\approx 0$.

Note, that sometimes geometry can change this behavior. For a 1D string of equidistant
emitters illumination at a chosen angle of incidence and polarization leads to a designable
phase gradient of the excitation amplitudes. The situation becomes even more complex for a 3D
cubic lattice, where the phases also differ in the different lattice planes. As a lucky coincidence,
a perpendicular plane illumination at the clock frequency in a magic lattice for Strontium (Sr) targets an
almost dark state. This leads to subradiance and in principle allows for a spectral resolution
better than the natural linewidth~\cite{maier2014superradiant}. In not so favorable cases
one could also think of a specific lattice design to facilitate a tailored dark state excitation.

\noindent \textbf{Radiative properties}. In order to be useful
resources for quantum information applications, target states should
exhibit \textit{robustness} with respect to the environmental
decoherence. To identify states of minimum decay rate, we scan
through the eigenstates ${\ket{\psi_k}}$ of the Hamiltonian
$H=H_0+H_{dip}$ (for $k=1,...,2^N$) and compute their decay rates
$\Gamma_{\psi_k}$ (see section Methods below). We find that
generally, for a given manifold, the energetic ranking of the states
closely indicates their robustness to decay (as illustrated by the
color-coding in Fig.~\ref{fig1}\textcolor{blue}{c}) ranging from
blue for subradiant states to red for superradiant states. This is
due to the fact that both radiation and energetic shifts are
strongly dependent on the symmetry of the states. In
Fig.~\ref{fig1}\textcolor{blue}{d}, for $N=6$, we plot the decay
rates of the collective states in the first ($n=1$) and second
($n=2$) excitation manifold arranged as a function of their
increasing energy corresponding to the level structure of
Fig.~\ref{fig1}\textcolor{blue}{c}. Superradiant states are found at
the upper sides of the manifolds while the ideal robust states lie
at the bottom. In Fig.~\ref{fig1}\textcolor{blue}{d}, the arrows
indicate the optimal decay rates in the single- ($0.0009\,\Gamma$)
and double-excitation manifolds ($0.0402\,\Gamma$) corresponding to
target states A and B whose population evolution is depicted in
Fig.~\ref{fig1}\textcolor{blue}{e}.

\noindent Within the single-excitation manifold, an analytical
expression for the decay rate of a state $\ket{m}$ can be found as
$\Gamma_m=\sum_{i,j}\gamma_{ij}f_i^mf_j^m$. For small distances the
state $m=1$ (upper state) is superradiant, whereas states at the
bottom of the manifold $m\sim N$ exhibit subradiant properties. In
the Dicke limit where $a=0$ we have $\gamma_{ij}=\Gamma$ for any $i$
and $j$, and we can compute $\Gamma_m=2\Gamma
\cot^2{[m\pi/(2N+2)]}/(N+1)$ for $m$ odd and $\Gamma_m=0$ for $m$
even. Note, that in this particular limit, these are the same
conditions as for the darkness and brightness of a state. For large
numbers of emitters, we recover the expected superradiant scaling
with $N$ for the state with $m=1$, i.e. $\Gamma_1\approx 8\Gamma
N/\pi^2$. On the other hand, large $m$ yield a decay rate of
$\Gamma_m\approx 0$ (perfect subradiance) in the same limit.

There are two important conclusions from these results: i) since in
the considered limit the decay rate of the superradiant state
$\ket{m=1}$ scales with $\Gamma_1\propto N$, whereas its driving is
$\chi_1\propto\sqrt{N}$, driving this state becomes more difficult
with increasing atom number due to the reduced time-scale and ii) if
the number of atoms is not too large, $\chi_m$ will remain finite,
while $\Gamma_m$ already indicates vast subradiance due to its
scaling-down with $N$. Hence, there are robust states that remain
bright, i.e. they can be driven directly even though the driving is
not matched to their symmetry.

\subsection*{Accessing dark states via magnetic field gradients}

\begin{wrapfigure}{r}{.5\textwidth}
\begin{center}
\includegraphics[width=0.5\textwidth]{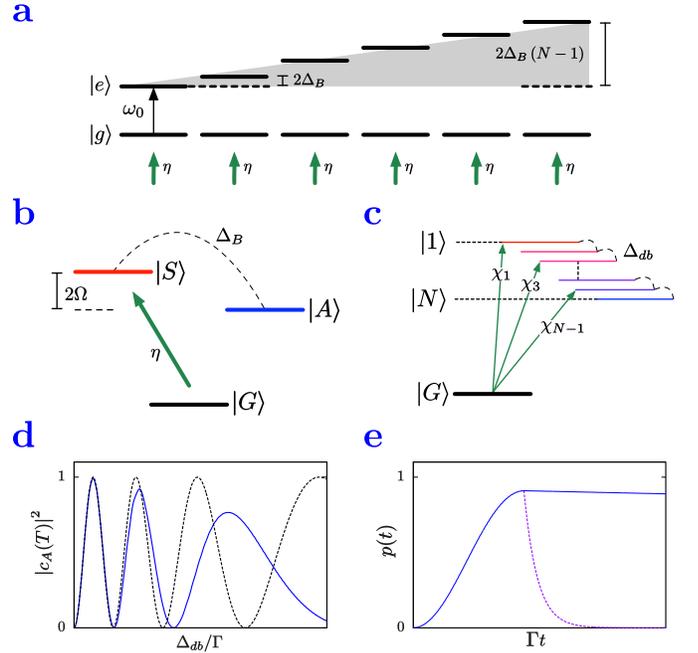}
\end{center}
\caption{\emph{Coupling to dark states via a magnetic field
gradient}. \textcolor{blue}{a)} Linearly increasing level shifts
along the chain occuring in the presence of the magnetic field
gradient. \textcolor{blue}{b)} Illustration of the level structure
and indirect dark state access for two coupled emitters. While
symmetry selects the state $\ket{S}$, off-resonant addressing
combined with bright-dark state coupling of strength $\Delta_B$
allows for a near-unity population transfer into the state
$\ket{A}$. \textcolor{blue}{c)} Dynamics in the single-excitation
manifold of $N$ coupled emitters where symmetric driving reaches the
bright states with amplitudes $\chi_m$ while the magnetic field
couples neighboring dark and bright states. \textcolor{blue}{d)}
Plot of the asymmetric state population for the two-atom case as a
function of the increasing magnetic field (solid line) compared to
the steady-state approximation (dashed line) at numerically
optimized time $T= 16.19\,\Gamma^{-1}$, with parameters
$\eta=\,\Gamma$ and $a=0.05\,\lambda_0$. \textcolor{blue}{e)} For a
chain of $N=4$ emitters, a $91\%$-efficient $\pi$-pulse to the most robust
state can be achieved as demonstrated in the population evolution
plot. The separation is chosen to be $a=0.025\,\lambda_0$, while
$\eta=40\,\Gamma$ and numerical optimization is employed to find
$\Delta_B=0.98\,\Gamma$.} \label{fig2}
\end{wrapfigure}

The direct symmetric driving with $H_{sym}$ allows access to bright
states only. Given that nearby dark states can conceivably be more
robust, we now employ a progressive level shifting mechanism
that allows for a coupling between bright and dark states. This is
achieved by subjecting the ensemble to a magnetic field with a
positive spatial gradient along the chain's direction. The
increasing energy shift of the upper atomic levels (as depicted in
Fig.~\ref{fig2}\textcolor{blue}{a}) plays a role similar to the
individual phase imprinting mechanism described previously. For each
particle the shift of the excited level induces a time-dependent
phase proportional to the value of the magnetic field at its
position. We demonstrate the mechanism for a particular two-atom
example, where indirect near unity access to the dark subradiant
asymmetric collective state is proven and extend it to the
single-excitation manifold of $N$ atoms.

\noindent\textbf{Two-atom case}. The eigenstates of the Hamiltonian
$H_0+H_{dip}$ are $\ket{E}=\ket{ee}$, $\ket{G}=\ket{gg}$ and in the
single-excitation subspace $\ket{S}=(\ket{eg}+\ket{ge})/\sqrt{2}$
and $\ket{A}=(\ket{eg}-\ket{ge})/\sqrt{2}$. The symmetric state
$\ket{S}$ is superradiant ($\Gamma_S=\Gamma_1=\Gamma+\gamma_{12}$)
and bright, directly accessible via symmetric driving with strength
$\chi_1=\sqrt{2}\eta$. The asymmetric state $\ket{A}$, on the other
hand, is subradiant ($\Gamma_A=\Gamma_2=\Gamma-\gamma_{12}$) and
dark. Indirect access can be achieved by shifting the second atom's
excited state by $2\Delta_B$ (see schematics in Fig.
\ref{fig2}\textcolor{blue}{b}), where $\Delta_B$ is tunable and
quantifies the per-emitter shift for a given magnetic field
amplitude. We first analyze the dynamics in the absence of decay by
solving the time-dependent Schr\"odinger equation governed by the
Hamiltonian $H=H_0+H_{dip}+H_{sym}+H_B$, where
$H_B=2\Delta_B\sigma_2^+\sigma_2^-$. We reduce the dynamics to three
states, and assume a quasi-resonant Raman-like scheme where the
population of $\ket{E}$ is at all times negligible. An effective
two-level system arises (between the ground state and the asymmetric
state; see section Methods below) and the resonance condition can be
identified as
\begin{align} \label{two-atom_resonance}
\Delta^{(2)} = -\Delta_B + \sqrt{\Delta_B^2+\Omega^2-2\eta^2},
\end{align}
with an effective Rabi frequency of
\begin{align} \label{two-atom_rabi}
\nu^{(2)}_R = \frac{\sqrt{2}\eta\Delta_B}{\Omega+\sqrt{\Delta_B^2+\Omega^2-2\eta^2}}.
\end{align}
To fulfill $|c_S|^2\ll1$, we need to restrict the
driving to a parameter regime where $\eta,\Delta_B\ll\Omega$. A scan over
the magnetic field is performed and the exact numerical results for the
asymmetric state population are plotted in Fig.~\ref{fig2}\textcolor{blue}{d}
against the adiabatic solution showing near unity population transfer for an
optimized $\Delta_B$. Further restrictions are imposed when decay is considered.
These stem from the fact that the coherent process described by $\nu_R$ should be
faster than the incoherent one characterized by $\Gamma_A$. For close particles,
the ability to tune the distance ensures that the scaling down of $\Gamma_A$ is very
fast and the above conditions are readily fulfilled. For the particular example
illustrated in Fig.~\ref{fig2}\textcolor{blue}{d} we chose $a=0.05\,\lambda_0$, resulting
in $\Omega=23.08\,\Gamma$, $\Gamma_A= 0.019\,\Gamma$. The $0.994$ population is reached
at $T= 16.19\,\Gamma^{-1}$, which is very close to the theoretical estimate
of $T=\pi/2\nu_R^{(2)}= 16.179\,\Gamma^{-1}$ obtained from the adiabatic
solution under the assumption of a $\pi$-pulse transferring the population to the target state.

\noindent \textbf{Many-atom case}. For a chain of $N$ atoms, we
consider the progressive shifting of excited levels along the chain
depicted in Fig.~\ref{fig2}\textcolor{blue}{a}. This is realized by
the application of a magnetic field with a constant gradient and is
described by the Hamiltonian
$H_B=2\Delta_B\sum_i(i-1)\sigma_i^+\sigma_i^-$. Let us consider a
dark state $\ket{d}$ ($d$ even) and the bright state $\ket{b=d-1}$
immediately above. Their coupling via $H_B$ is quantified by
$\Delta_{db}=2\Delta_B\sum_i(i-1)f_i^d f_i^{b}$, as shown in Fig.
\ref{fig2}\textcolor{blue}{c}.

\noindent We develop a protocol where direct off-resonant driving
into the bright state (amplitude $\chi_b$) combined with a coupling
between the bright and dark states via the magnetic field leads to
an almost unity population transfer into the dark state. Given a
sufficient energy separation, the problem can be reduced to solving
the time-dependent Schr\"odinger equation for the three coupled
state amplitudes $c_b, c_d$ and $c_G$. Following the same adiabatic
approximation as in the two-atom case we reduce the general dynamics
to an effective two-level system between the states meant to be
connected by an effective $\pi$-pulse, i.e. $\ket{d}$ and $\ket{G}$. The
generalized resonance condition (with
$\epsilon_{db}=\epsilon_d-\epsilon_b$) reads
\begin{align} \label{many-atom_resonance}
\Delta^{(N)} &=-\Delta_B(N-1)-\frac{\epsilon_d+\epsilon_{b}}{2} +\sqrt{\frac{\epsilon_{db}^2}{4}+\Delta_{db}^2-\chi_{b}^2},
\end{align}
and was obtained in the limit where the coupling of the dark state
to the other adjacent bright state $\ket{d+1}$ was neglected owing
to the relation $\chi_{d-1}\gg\chi_{d+1}$. The effective transition
rate between the ground state and the state $\ket{d}$ is
\begin{align} \label{many-atom_rabi}
\nu^{(N)}_R=\frac{\chi_{b}|\Delta_{db}|}{\Delta+\epsilon_{b}+\Delta_B(N-1)}.
\end{align}
The addition of decay imposes a new constraint on the timescale of
the process, i.e. $\nu^{(N)}_R\gg\Gamma_d$, required to ensure near
unity population in the dark state. The fulfillment of this
condition depends on the individual system under consideration. As
an illustration of the procedure, Fig. \ref{fig2}\textcolor{blue}{e}
presents the targeting of a robust dark state in the single
excitation manifold of four particles. Note, that the numerical
results are performed in an exact regime beyond the NN approximation
and are in excellent agreement with our conclusions obtained from
the NN treatment.

\section*{Discussions}

\subsection*{Entanglement properties}

To justify the usefulness of collective states for quantum information purposes, we employ the von Neumann entropy to analyze their entanglement properties.
More specifically, we compute the von Neumann entropy of the reduced
density matrix $\rho_s$ of a single two-level emitter (showing the
degree of its bipartite entanglement with the rest of the system)
defined by $S(\rho_s)=-\sum_i\lambda_i\log_2{\lambda_i}$, where
$\lambda_i$ is the $i$-th eigenvalue of $\rho_s$ and
$0\log_2{0}\equiv 0$. We furthermore minimize the set of
\begin{wrapfigure}{r}{.65\textwidth}
\begin{center}
\includegraphics[width=0.65\textwidth]{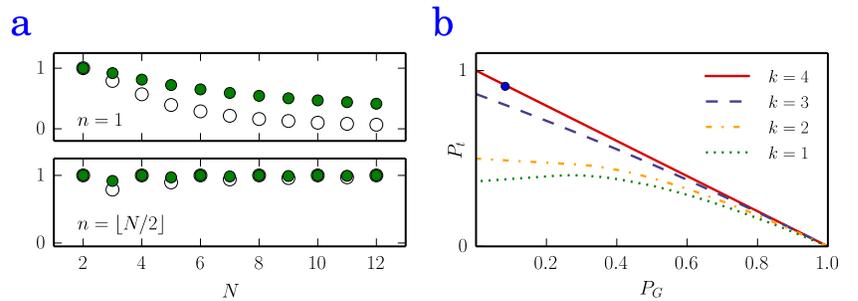}
\end{center}
\caption{\textit{Entanglement properties}. \textcolor{blue}{a)}
Comparison of the numerically computed von Neumann entropy (empty
circles) of the reduced density matrix of the chain minimized over
the atom index and the analytical expression for the entropy of the
Dicke state (green circles), both for excitations $n=1$ and
$n=\lfloor N/2\rfloor$ as a function of the atom number $N$ at distance $a=0.1\lambda_0$.
\textcolor{blue}{b)} Depth of entanglement of the subradiant
four-atom state (blue dot) prepared by the magnetic field gradient scheme (see
Fig. \ref{fig2}\textcolor{blue}{e}). It clearly
lies above the $k=3$ boundary indicating four-atom entanglement. The
$k$-atom entanglement boundaries of the target state population
$P_t$ as a function of the ground state population $P_G$ have been
computed for the corresponding target state of a four-atom chain at
distance $a=0.025 \,\lambda_0$.} \label{fig3}
\end{wrapfigure}
values for all atoms to obtain a lower bound on the entanglement
contained in the system. We compare the numerical results to the
single-atom entropy of the symmetric Dicke state $\ket{-N/2,-N/2+n}$
\cite{dicke1954coherence}. For these particular states the entropy
is maximized if the number of excitations in the state is $n=N/2$.
It follows that it is highly desirable to drive the system into
robust states as close as possible to $n=\lfloor N/2\rfloor$
excitations (where $\lfloor N/2\rfloor$ is the largest integer
smaller or equal to $N/2$), since this manifold contains the most
entangled state. A comparison of the exact numerical data and the
analytical expression for the entropy is shown in
Fig.~\ref{fig3}\textcolor{blue}{a}.

Another way to characterize the entanglement of the prepared state
is to investigate their \textit{depth of
entanglement}\cite{sorensen2001entanglement,haas2014entangled},
which does not quantify the entanglement itself but rather shows how
many atoms of an ensemble are involved in the present entanglement.
This measure has been used in recent experiments
\cite{haas2014entangled,mcconnell2015entanglement} since it is a
readily measurable quantity. The depth of entanglement is computed
as follows: given an $N$-atom target state in which an arbitrary
number of said $N$ atoms is entangled, we compute the limit of how
much population one can drive into this state such that the
resulting density matrix $\rho$ remains separable into a subset of
density matrices that exhibit no more than $k$-atom entanglement
($1\leq k\leq N$). This may be done by numerically maximizing the
target state population $P_t$ as a function of the ground state
population $P_G$ for different $k$. The boundaries themselves
indicate how many atoms need to be entangled in order to prepare the
pure target state, i.e. the boundary where the target state
population is maximized to $1$ corresponds to the number of atoms
entangled in the (pure) target state. If a general prepared state
has a target and ground state population such that the corresponding
data point lies on or above the $k$-atom boundary, more than $k$
atoms are entangled.

Obviously, for the pure target states considered in the above
computation all atoms contribute to the entanglement, since
otherwise the minimal von Neumann entropy as shown in
Fig.~\ref{fig3}\textcolor{blue}{a} would be zero. For a more
interesting result, we can compute the depth of entanglement in
order to demonstrate the efficiency of the driving procedure using a
magnetic field gradient as in Fig.~\ref{fig2}\textcolor{blue}{e}. From
Fig.~\ref{fig3}\textcolor{blue}{b}, where all boundaries have been plotted for the
considered subradiant four-atom state, it is clear that the prepared
state shows all-atom entanglement as the corresponding data point lies far above the boundary for three-atom entanglement.

\subsection*{Implementation considerations}

The proof-of-principle technique presented above has been particularized
on a specific generic system of emitters in an equidistant chain.
The choice is natural since the electromagnetic vacuum provides a
simple example for both collective dispersive and dissipative
dynamics. To exemplify a possible realization we consider a
particular system~\cite{olmos2013long} where bosonic Sr
atoms are trapped in a magic wavelength optical lattice at
separations of $a=206.4$nm. The working transition is at
$\lambda_0=2.6 \mu$m, between the $^3P_0$ and $^3D_1$ electronic
states. This amounts to a ratio of $a/\lambda_0 \approx 1/13$ which
allows for an operation in the regime targeted by our scheme. The
corresponding single atom decay rate is at the order of $\Gamma=0.3$
MHz and circularly polarized light can allow for transitions between
states with a difference of 1 in magnetic quantum number. We have
numerically investigated a system of 4 atoms in such a configuration
and found a sizeable $73\%$ target state population for
$\eta=2\,\Gamma$ and $\Delta_B=0.5\,\Gamma$, under the conditions of
a relatively small level shift between the dark and bright state
around $6\,\Gamma$ which does not allow for large driving powers.
For further optimization of the efficiency of the target state
preparation one could envision a modified setup where a trapping
transition of smaller wavelength can be chosen that would most
importantly allow for better state separation (owing to larger
dipole shifts). The corresponding magnetic field gradient required
to produce the considerable $\Delta_B=0.5\,\Gamma$ shift on a
distance of $a=206$ nm is around $5.2\cdot 10^5$ G/m, not far from
state-of-the-art values achievable in high magnetic field gradient
magneto-optical trap
experiments~\cite{ueberholz2002cold,yoon2007characteristics}. Of
course, there are many detrimental practical effects that can
seriously limit the above technique such as light-assisted collision
loss. We envision the extension of the described technique to
systems where both the coherent and dissipative particle-particle
interactions can be suitably tailored. For example, the same kind of
dipole-dipole Hamiltonians can occur in 3D lattices of polar
molecules~\cite{yan2013observation} or between two different color
NV centers in diamonds~\cite{neumann2010quantum}.

\subsection*{Conclusions}

Direct particle interactions are typically detrimental and limiting in precision measurement
applications. Here, we have presented some specific opposite
examples, where the \textit{collective} nature of the decoherence
combined with the coherent binary dipole-dipole interactions is used
as a new resource for the controlled and efficient preparation of
specially selected states. The excitation scheme can be tailored to
address target states exhibiting both entanglement as well as
robustness against decay. As a generic example we studied the case
of a one-dimensional system of tightly spaced equidistant quantum
emitters. Already the inherent dipole-dipole coupling allows for a
targeted state preparation technique via energy selection. The
performance of the excitation can be enhanced additionally via the
\textit{continuous} application of a spatially increasing magnetic
field. The general principle of such a phase imprinting technique is
potentially applicable in many specific environments such as optical
lattices or atoms and ions localized within one or more common
optical cavity modes~\cite{zoubi2010hybrid,meiser2010steady},
NV-centers or superconducting qubits coupled to CPW transmission
lines or resonators~\cite{Sandner2010strong,lalumiere2013input}.

\section*{Methods}

\subsection*{Decay rate of the states}

In order to arrive at an analytical expression for the decay rate of
an eigenstate $\ket{\psi_k}$ of the Hamiltonian in
equation~\eqref{H}, we consider the homogeneous part of the
differential equation of the corresponding density matrix element
that arises from the master equation. The solution of this
differential equation yields an exponential decay. The rate at which
the state population decays may be written as
\begin{align} \label{Gamma_psi}
\Gamma_{\psi_k}=-\bra{\psi_k}\mathcal{L}\left[\ket{\psi_k}\bra{\psi_k}\right]\ket{\psi_k} = \sum_{i,j}\gamma_{ij}\bra{\psi_k}\sigma_i^+\sigma_j^-\ket{\psi_k}.
\end{align}
Note, that this is true only for states that contain one specific
number of excitations, i.e. they are eigenstates of the operator
$\sum_i\sigma_i^z$. Obviously, this is fulfilled for eigenstates of
the considered Hamiltonian. Equation~\eqref{Gamma_psi} was used in
order to compute the rates depicted in Fig.
\ref{fig1}\textcolor{blue}{d} and throughout the manuscript. For
example, we used it in order to compute the decay rate of the
eigenstates in the NN approximation $\Gamma_m$.

\subsection*{Subradiance and disorder}
\begin{wrapfigure}{r}{.65\textwidth}
\begin{center}
\includegraphics[width=0.65\textwidth]{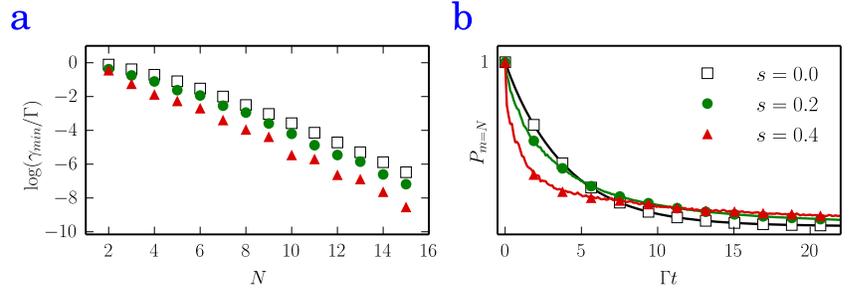}
\end{center}
\caption{\emph{Subradiance and disorder.} a) Plot of the logarithm of the minimal eigenvalue of the decay rate matrix (matrix with entries $\gamma_{ij}$)
as a function of $N$ at a distance of $a=0.4\lambda_0$ for increasing levels of disorder ($s=0,0.2,0.4$).
b) Decay of the $\ket{m=N}$ state as a function of time. In the presence of disorder ($s=0.2,0.4$) the short time and long time behaviors are fundamentally different. At short times,
disorder can push the state towards faster decaying channels while decay inhibition due to disorder occurs at larger times.} \label{fig4}
\end{wrapfigure}

Let us consider the influence of positioning disorder on subradiant
properties of the target states. To mimic disorder we perturb an
equidistant chain of $N$ emitters (average separation $a$) by
introducing an uncertainty in each emitter position quantified by a
defect parameter $s$ (normal distribution of variance $sa$). We then
write the randomized matrix of decay rates and find the minimum
decay channel without as well as in the presence of disorder of
$s=20\%$ and $s=40\%$. For the $s=0\%$ case, it has been
shown\cite{ostermann2014protected} that the minimum decay rate
scales exponentially with $N$ even for distances up to
$0.4\lambda_0$, while the linear scaling with $N$ typical for
superradiance is reached for $a\ll\lambda_0$ only. After averaging
over $100$ random configurations, we plot the logarithm of the
minimal rates as a function of increasing $N$ in
Fig.~\ref{fig4}\textcolor{blue}{a}.
As a somewhat surprising result,
subradiance scales even better with $N$ as the disorder increases.
This might be understood as a destructive interference effect
brought on by the cancelation of emitted photons stemming from the
random positioning. As pointed out in previous
investigations~\cite{ostermann2014protected}, the states of low
symmetry (as, for example, the $m=N$ state) possess decay rates
closest to the analytically derived minimal rate. We analyze the
respective sensitivity of the state subradiance to disorder by
initializing the system of $N$ emitters in the $m=N$ state and allow
it to decay. The outcome is plotted in
Fig.~\ref{fig4}\textcolor{blue}{b} and shows remarkable robustness
of the disordered systems on a long time-scale. While on a short
time-scale disorder pushes the considered state into faster decaying
channels, the long time limit shows that the remaining population
accumulates in the disorder-enhanced robust states.

For short time-scales, the state still
decays slowly (subradiantly), however, the decay rate increases with growing disorder ($s=40\%$).
More remarkable, though, is the behavior the decaying states show for long time-scales, as the states
subject to larger disorder become more robust than the unperturbed system. This is due to the fact that
all population in the $m=N$ state that decays through more radiative channels have decayed at that point
and only the most subradiant channel (minimal eigenvalue of the decay rate matrix) remains. As seen in Fig.~\ref{fig4}\textcolor{blue}{a},
this eigenvalue is even further reduced by disorder which explains the long time-scale behavior in Fig.~\ref{fig4}\textcolor{blue}{b}.

\subsection*{Coherent dynamics with a magnetic field gradient}
\textbf{Two-atom case}. To find the expressions in
equation~\eqref{two-atom_resonance} and
equation~\eqref{two-atom_rabi} we solve three coupled differential
equations neglecting the population of the fully inverted state
$\ket{E}$ as far off-resonant for all times. In the collective
basis, where any state may then be written as $\ket{\psi} = c_S
\ket{S} + c_A \ket{A} + c_G \ket{G}$, the equations are
\begin{align}
i\dot{c}_S &= (\Delta+\Delta_B+\Omega)c_S - \Delta_B c_A + \sqrt{2}\eta c_G,
\\
i\dot{c}_A &= (\Delta+\Delta_B-\Omega)c_A - \Delta_B c_S,
\\
i\dot{c}_G &= \eta c_S,
\end{align}
where $\Omega=\Omega_{12}$ is the coherent
interaction between the atoms and $\Delta$ is the detuning between
the atomic resonance frequency and the driving laser. For an efficient driving of $\ket{A}$ the
population of the state $\ket{S}$ needs to be negligible which allows
us to set a steady-state condition, namely $\dot{c}_S=0$ yielding the desired effective two-level system between $\ket{G}$ and $\ket{A}$.

\noindent \textbf{Many-atom case}. The same approach as in the
two-atom case may be used to describe the dynamics in the
single-excitation manifold for an arbitrary number of atoms in a
chain. Given sufficient energy separation we may neglect all states
but the ones we aim to address. We can indirectly address a dark
state $\ket{d}$ by driving the bright state $\ket{b}$ immediately
above, which is coupled to the dark state by a magnetic field
gradient. Neglecting all populations but $c_b$, $c_d$, and $c_G$ and
their respective couplings via the magnetic field gradient, the
investigation reduces to the equations
\begin{align}
i\dot{c}_b&=\left[\Delta+\epsilon_b+\Delta_B(N-1)\right]c_b+\Delta_{db}c_d+\chi_bc_G,
\\
i\dot{c}_d&=\left[\Delta+\epsilon_d+\Delta_B(N-1)\right]c_d+\Delta_{db}c_b,
\\
i\dot{c}_G&=\chi_bc_b.
\end{align}
For an efficient driving of the dark state we may again invoke a
steady-state condition on the bright state population $\dot{c}_b=0$.
This, again, yields an effective two-level system between the ground
and the dark state with resonance condition and Rabi frequency as
displayed in equation~\eqref{many-atom_resonance} and
equation~\eqref{many-atom_rabi}, respectively.

\subsection*{Von Neumann entropy}

For a Dicke state an analytical expression for the von Neumann entropy of
the reduced density matrix can be obtained. First, note that, since Dicke states are
invariant under a permutation of the atoms, all reduced density matrices are identical.
Hence, they all share the same von Neumann entropy for a given number of excitations $n$.
We may choose to reduce the full density operator $\rho$ to the density matrix of the first
atom in the ensemble, i.e. $\rho_s^1 \equiv \rho_s = \text{tr}_{2,...,N}(\rho)$ which yields a von Neumann entropy of
\begin{align}
S(\rho_s) &= \frac{n}{N}\log_2{\left(\frac{N}{n}\right)} - \left(1-\frac{n}{N}\right)\log_2{\left(1-\frac{n}{N}\right)}.
\end{align}
For the actual eigenstates of the Hamiltonian in equation~\eqref{H}
this computation needs to be done numerically. Furthermore, these
states are not invariant under permutation of atoms and hence it is
required to minimize the entropy with respect to the atomic chain
index in order to find the lower bound.

\subsection*{Depth of entanglement}

The boundaries depicted in Fig. \ref{fig3}\textcolor{blue}{b} were
found by maximizing the target state population with the condition
on the density matrix of the prepared state to contain no more than
$k$-atom entanglement, i.e. $\rho=\bigotimes_i\rho_i^{k_i}$ with
$k_i\leq k$ and at least one $k_i=k$. To compute the boundaries we
generalized the algorithm that was previously used solely for the
$W$-state~\cite{haas2014entangled} to arbitrary states in the
single-excitation manifold. For the computation of all boundaries we
need to distinguish the two cases where $P_G=0$ and $P_G>0$.
Considering a separable state ($k=1$), the boundary for $P_G>0$ is
found to be
\begin{align}
\max(P_t) &= P_G \max_{\prod_i\alpha_i=\sqrt{P_G}}\left|\sum_i|c_i|\frac{\sqrt{1-\alpha_i^2}}{\alpha_i}\right|^2,
\end{align}
where $\alpha_i~\in~[0,1]$ and $c_i$ are the coefficients of the
target state. For $P_G=0$ the maximization is much simpler, i.e.
$\max(P_t)=\max{|c_i|}^2$, which is found by setting one
$\alpha_i=0$ and the remaining coefficients $\alpha_{j\neq i}=1$.
Note, that for both these and all following computations we neglect
the symmetry of the state, i.e. the phases of the coefficients $c_i$
by using $|c_i|$. This is valid due to the invariance of
entanglement under local unitary operations and necessary if we
restrict the coefficients $\alpha_i$ in the way we did.

For multiple-atom entanglement ($k>1$) the matter of finding the
corresponding boundary is no longer so simple. In order to find the
maximum population, we assume maximally allowed entanglement in the
prepared state. We split the prepared state into $M=\lceil
N/k\rceil$ sets, where $M-1$ sets are $k$-atom entangled and the
remaining one is $k'=N-k(M-1)$-atom entangled. To find the maximum,
one has to consider all possible positions of the $k'$-entangled
state. If, for example, the $k'$-entangled state is at the last
position, the population of the target state $\ket{t}$ in the
prepared state reads
\begin{align}
P_t &= \left|\bra{t}\left[\left(\bigotimes_{i=1}^{M-1}\ket{\varphi_i^k}\right)\otimes\ket{\varphi_M^{k'}}\right]\right|^2,
\end{align}
where
\begin{align}
\ket{\varphi_i^k}=\alpha_i\ket{G_k}+\sqrt{1-\alpha_i^2}\sum_{r=1}^k\lambda_r^i\sigma_r^+\ket{G_k}
\end{align}
is a general non-separable state of $k$ atoms in the
single-excitation manifold. The state $\ket{G_k}$ is the $k$-atom
ground state and the coefficients $\lambda_r^i~\in~[0,1]$ have to be
normalized, i.e. $\sum_r(\lambda_r^i)^2=1~\forall~i$. One then has
to maximize the target state population with respect to the
coefficients $\alpha_i~\in~[0,1]$ and $\lambda_r^i$ with the
condition $\prod_j\alpha_j=\sqrt{P_G}$. The number of these
coefficients, however, grows vastly with the number of atoms, hence
numerical computations are limited. For $P_G=0$ one can again choose
one $\alpha_i=0$ and all $\alpha_{j\neq i}=1$.

Note, that all boundaries computed via this maximization only hold
for pure states. In order to find the boundaries for mixed states we
need to compute the convex hulls of the respective
boundaries~\cite{haas2014entangled}. The $k=N$ boundary is found
when a perfect superposition between the ground and target state is
reached.

In this work we considered the specific case of an exciton state of
a four-atom chain. In that case, when investigating two-atom
entanglement the permutation of the $k'$-entangled state is rendered
unnecessary since $k'=k=2$. Unfortunately, this is no longer true
for $k=3$, where we did have to account for all permutations.


\section*{Acknowledgements}

We acknowledge financial support from the SFB through the FoQus project (D.~P.), DARPA through the QUASAR project (L.~O. and H.~R.) and from the Austrian Science Fund (FWF) via project P24968-N27 (C.~G.). Furthermore, we acknowledge the use of the QuTip open-source software \cite{johansson2013qutip}. H.R. thanks Vladan Vuletic for helpful discussions. C.~G. thanks M.~W.~Mitchell for the suggestion of energetic addressing of collective states.

\section*{Author contributions statement}

C.~G. conceived the ideas and supervised the work. D.~P. developed the concepts, conducted analytical calculations, took the main role in writing the manuscript and wrote numerical simulations, with support from L.~O. especially in generalizing the depth of entanglement. H.~R. provided guidance and expertise. The manuscript has been reviewed and edited by all authors.

\section*{Additional information}

No competing financial interests arise in the funding of this work.

\end{document}